\documentclass[pdflatex,sn-mathphys-num]{sn-jnl}

\usepackage{graphicx}%
\usepackage{multirow}%
\usepackage{amsmath,amssymb,amsfonts}%
\usepackage{amsthm}%
\usepackage{mathrsfs}%
\usepackage[title]{appendix}%
\usepackage{xcolor}%
\usepackage{textcomp}%
\usepackage{manyfoot}%
\usepackage{booktabs}%
\usepackage{algorithm}%
\usepackage{algorithmicx}%
\usepackage{algpseudocode}%
\usepackage{listings}%
\usepackage{makecell}
\usepackage{multirow}
\usepackage{subcaption}
\usepackage{ragged2e}
\usepackage{comment}

\theoremstyle{thmstyleone}%
\theoremstyle{thmstyletwo}%
\newtheorem{remark}{Remark}%

\theoremstyle{thmstylethree}%

\begin{document}
\newcommand{\R}{{\mathbf{R}}}
\newcommand{\norm}[1]{\left\lVert#1\right\rVert}

\title[Article Title]{ Distributed Pose Graph Optimization using the Splitting Method based on the Alternating Direction Method of Multipliers}


\author*[1]{Zeinab Ebrahimi}\email{z.ebrahimi@unsw.edu.au}

\author[1]{Mohammad Deghat}\email{m.deghat@unsw.edu.au}

\affil*[1]{\orgdiv{Mechanical and Manufacturing Engineering}, \orgname{University of New South Wales}, \orgaddress{ \city{Sydney}, \postcode{2016}, \state{NSW}, \country{Australia}}}


\abstract{Distributed optimization aims to leverage the local computation and communication capabilities of each agent to achieve a desired global objective. This paper addresses the distributed pose graph optimization (PGO) problem under non-convex constraints, with the goal of approximating the rotation and translation of each pose given relevant noisy measurements. To achieve this goal, the splitting method based on the concepts of the alternating direction method of multipliers (ADMM) and Bregman iteration are applied to solve the rotation subproblems. The proposed approach enables the iterative resolution of constrained problems, achieved through solving unconstrained problems and orthogonality-constrained quadratic problems that have analytical solutions. The performance of the proposed algorithm is compared against two practical methods in pose graph optimization: the Distributed Gauss-Seidel (DGS) algorithm and the centralized pose graph optimizer with an optimality certificate (SE-Sync). The efficiency of the proposed method is verified through its application to several simulated and real-world pose graph datasets. Unlike the DGS method, our approach attempts to solve distributed PGO problems without relaxing the non-convex constraints.}




\maketitle

\section{Introduction}\label{sec1}

In recent decades, the rapid evolution of communication, sensing, and wireless technologies has fostered the growth of networked systems, where multiple interconnected agents collaborate to achieve global objectives. Optimization algorithms for networked systems can be categorized as either centralized or distributed. Conventional centralized schemes face challenges in solving significant networked optimization problems due to their high communication and computation demands, as well as limited scalability. To overcome these limitations, researchers have turned their attention to distributed optimization techniques, which have found applications in various fields, including smart grids, wireless sensor networks, robotics, and machine learning.

Distributed optimization entails the endeavor of minimizing a collective objective function, which comprises a summation of various localized objective functions, each associated with an individual computational node. While distributed optimization has been a subject of extensive investigation in the optimization community \cite{rockafellar1976monotone, tsitsiklis1984problems}, its integration into the realm of robotics has been observed in only a limited number of instances. Nonetheless, the employment of distributed optimization techniques bears significant implications for multi-robot systems, as numerous pivotal tasks within this domain, such as cooperative estimation \cite{shorinwa2020distributed}, multi-agent learning \cite{wai2018multi}, pose graph optimization \cite{thrun2008simultaneous}, and collaborative motion planning \cite{bento2013message}, can be framed as distributed optimization problems. The utilization of a distributed optimization framework presents a versatile and potent approach for devising effective and decentralized algorithms catering to a multitude of multi-robot predicaments.

Among the fundamental non-convex optimization problems, Pose Graph Optimization (PGO) plays a crucial role in applications like sensor network localization, camera motion/orientation estimation in computer vision, and distributed consensus optimization on manifolds \cite{simonetto2014distributed, tron2012intrinsic,sarlette2009consensus}. In these scenarios, agents approximate their current states based on relative measurements among other agents. Additionally, PGO finds significant application in the context of the Simultaneous Localization and Mapping (SLAM) problem \cite{durrant2006simultaneous}, where robots must estimate their poses from noisy measurements and limited communication with other robots. However, traditional iterative PGO solvers, like the Gauss-Newton method and trust-region techniques, may only guarantee local convergence, failing to find the global minimum even with small noise variations \cite{kummerle2011g, rosen2014rise}. To enhance global convergence, a strategy involving rotation estimation followed by iterative solvers has been proposed \cite{carlone2014angular}.

Among the commonly used techniques for distributed PGO, the Alternating Direction Method of Multipliers (ADMM) stands out for its simplicity, minimal information exchange, and lack of linearization requirements \cite{choudhary2015exactly}. Although ADMM provides acceptable solutions, its convergence rate still falls short of state-of-the-art methods. To address this issue, simultaneous parallelization across robots or processing units is employed.

Despite the promising results demonstrated by current methods in distributed optimization, a gap exists between these approaches and centralized methods in accurately determining the global objective value in PGO problems. This work aims to bridge this gap by developing an efficient algorithm for distributed PGO problems with non-convex constraints that consider both translation and rotation estimation.

\subsection{Related Work}

Pose Graph Optimization (PGO) serves as a mechanism for estimating a collection of $\mathrm{SE}(3)$ poses, which arise from the interplay of noisy relative measurements among these poses. Characterized by a non-convex nature, PGO entails a least-squares optimization problem that pertains to rotation matrices and translation vectors linked to each pose within networked systems.

Recent advancements in PGO have led to notable contributions. Notably, Rosen \textit{et al}. \cite{rosen2019se} introduced a certifiably correct technique termed SE-Sync, premised on convex semi-definite relaxation. SE-Sync strives to attain an optimal global solution for PGO, provided that existing measurement noise remains below a specific threshold. However, the applicability of the SE-Sync algorithm within distributed implementations is hampered by the involvement of fully dense matrices in semi-definite relaxation. Furthermore, the centralized nature of this approach necessitates substantial communication among agents, particularly within networks comprising a large number of agents. These limitations have fueled the interest in exploring distributed approaches, wherein robots communicate within localized spheres to establish a consensus on trajectory estimation.

Chouldhary \textit{et al}. \cite{choudhary2017distributed} proposed a prominent two-step method for solving distributed PGO. This approach first obtains an unconstrained linear least-squares rotation problem by relaxing non-convex constraints. Subsequently, the obtained rotations initialize a second step involving a single Gauss-Newton iteration to address the PGO problem. The integration of Jacobi Over-Relaxation (JOR) and Successive Over-Relaxation (SOR) \cite{bertsekas2015parallel} as iterative linear solvers in both steps results in improved performance compared to preceding endeavors \cite{cunningham2013ddf}. Nevertheless, further exploration within non-convex settings is warranted, as results lack performance guarantees.
Tian \textit{et al}. \cite{tian2021distributed} presented a distributed certifiably correct PGO solver predicated on sparse semi-definite relaxation. This methodology leverages Riemannian block coordinate descent to manage the semi-definite relaxation problem. This approach ensures algorithm convergence to globally optimal solutions, even in the presence of measurement noise. Despite the integration of Nesterov's acceleration method, the practicality of \cite{tian2021distributed} diminishes when addressing extensive networked optimization problems without master nodes, which can lead to decrease in computational performance.

Exploration into distributed 3D pose estimation from noisy relative measurements remains limited. Tron \textit{et al}. \cite{tron2014distributed} addressed this through distributed gradient descent, encompassing convex relaxation for rotation, translation, and overall 3D pose objectives. Meanwhile, Thunberg \textit{et al}. \cite{thunberg2017distributed} tackled synchronization under SE(3), validating convergence via the projection of rotation approximations onto SO(3). Notably, these contributions lack consistent consideration of noisy measurements. Cristofalo \textit{et al}. \cite{cristofalo2019consensus} introduced minimum consistency conditions in measurements, particularly in rotation, translation, and 3D pose approximation, capitalizing on the axis-angle rotation model over an undirected graph. However, the exclusion of translation measurements and focus solely on pairwise consistent PGO represents a limitation.

In a parallel track, the Alternating Direction Method of Multipliers (ADMM) emerges as a heuristic approach for distributed PGO. Choudhary \textit{et al}. \cite{choudhary2015exactly} employed a separator method, splitting graphs into subgraphs to steer convergence, yet without guarantees. Bartali et al. \cite{bartali2024consensus} proposed a consensus-based ADMM framework for the distributed Minimum Lap-Time Problem (MLTP), focusing on optimizing vehicle trajectories over long racetracks. Their approach divides the track into segments, solving subproblems locally while ensuring global consistency through ADMM iterations. Despite advancements such as Nesterov acceleration \cite{bosland2021nesterov}, challenges remain in securing rapid convergence. Sebastián et al. \cite{sebastian2024accelerated} proposed the Accelerated ADMM Gradient Tracking (A2DMM-GT) algorithm, which integrates momentum acceleration with ADMM for unconstrained consensus optimization over static undirected networks. The proposed method in \cite{sebastian2024accelerated} could achieve faster convergence rates by leveraging the robustness of ADMM and the computational efficiency of gradient tracking, particularly in scenarios with smooth and strongly convex cost functions. While A2DMM-GT demonstrates significant advancements in distributed optimization, its focus on unconstrained problems limits its applicability to structured non-convex optimization problems. Noteworthy endeavors have been undertaken in employing ADMM for nonconvex optimization challenges, as evident in works such as \cite{yi2022sublinear, engelmann2020decomposition}. Despite these notable advancements, the domain of solving distributed nonconvex optimization problems confronts ongoing challenges in developing algorithms that are both efficient and scalable. This is particularly relevant within pose graph optimization settings. 
Overall, the landscape of distributed optimization within the PGO sphere necessitates continued exploration and innovation to address the intricacies posed by non-convexity and distributed implementation.

\subsection{Contribution}

In the PGO problem, the objective is a quadratic function. The complexity of the optimization problem stems from the fact that the constraint related to the rotation is non-convex. Currently, available distribution optimization algorithms rely on the approximation of rotation at first. Then the known rotation is employed for solving the least square problems for translation approximation. In contrast to the aforementioned methods, our technique addresses the rotation subproblem without relaxing the non-convex constraints. Afterward, the full poses would be recovered.     

In recent years, there has been a growing interest in constraint preserving algorithms, which are based on an investigation of the Stiefel manifold structures of orthogonality constraints. Notably, the most advanced approach to this area is presented in \cite{goldfarb2009curvilinear,wen2013feasible}, where a curvilinear search strategy is introduced using the Cayley transformation and Barzilai-Borwein step size selection. Although the feasible approaches employed in these methods are mathematically elegant, they rely on the manifold structure of the Stiefel manifold. Therefore, the splitting of the orthogonality constraints (SOC) method \cite{lai2014splitting} is adopted to handle the complicated manifold structures of the constraints. 

In this paper, the subproblem related to the rotation update is in the form of an orthogonal constraint problem. To address the non-convex subproblem, the SOC method employs a combination of the ADMM approach and the split Bregman iteration method. Through the iterative update of variables, this approach provides a solution to the rotation subproblems. Contrary to the existing state-of-the-art techniques \cite{choudhary2015exactly, choudhary2017distributed}, our proposed algorithm is both distributed and has a good error and convergence time owing to its underlying splitting procedures. Moreover, we illustrated that the whole procedure of solving the pose graph optimization problem is close to the global minima. 

\section{Background of Bregman Iteration and Its Equivalence to ADMM }
Bregman iteration, introduced by Osher \textit{et al}. \cite{osher2005iterative}, is an effective method for solving optimization problems involving total variation, particularly in image processing. Over the years, the Bregman iteration method has attracted significant attention due to its efficiency in solving constrained optimization problems, which can be formulated as follows \cite{esser2009applications, osher2005iterative}:
\begin{equation} \label{problem bregman}
   \min_x J(x), \quad \text{subject to } \mathcal{K}x = f, 
\end{equation}
where $x \in \mathbb{R}^m$, $f \in \mathbb{R}^s$, $\mathcal{K} \in \mathbb{R}^{s \times m}$, and $J(x)$ is a convex objective function.
\subsection{Bregman Iteration}
The Bregman iteration solves this problem through an iterative procedure involving the following updates:
\begin{align} \nonumber \label{update B}
    &x^{k+1} = \arg\min_x \mathcal{B}_{J}^{p^k}(x, x^k) + \frac{\alpha}{2} \| \mathcal{K}x - f \|_2^2 ,\\
    &p^{k+1} = p^k - \alpha \mathcal{K}^\top (\mathcal{K}x^{k+1} - f)
\end{align}
where $\alpha$ denotes the penalty parameter and $\mathcal{B}_{J}^{p^k}(x, x^k)$ is the Bregman distance between $x$ and $x^k$, defined as:
\begin{align}
    \mathcal{B}_{J}^{p^k}(x, x^k) = J(x) - J(x^k) - \langle p^k, x - x^k \rangle.
\end{align}
where $p^k$ is a subgradient of $J$ at $x^k$ for $k \geq 1$.
Yin \textit{et al.} \cite{yin2008bregman} has shown that the update procedure in (\ref{update B}) can also be represented as the following simplified two-step approach using a Bregman penalty function:
\begin{align} \nonumber \label{bregman update}
    &x^{k+1} = \arg\min_x J(x) + \frac{\alpha}{2} \| \mathcal{K}x - f + b^k \|_2^2,\\
    &b^{k+1} = b^k + \mathcal{K}x^{k+1} - f.
\end{align}

Yin \textit{et al.} \cite{yin2008bregman} demonstrates the equivalence between Bregman iteration and the method of multipliers for linear constraints by defining $\lambda^k$ for $k \geq 0$ and $\lambda^0=0$ as follows:
\begin{equation}
    \lambda^{k+1} = \lambda^k - \alpha ( \mathcal{K}x^{k+1} -f)
\end{equation}
Take into account that if $p^k=\mathcal{K}^\top \lambda^k$, then $p^{k+1}=\mathcal{K}^\top \lambda^{k+1}$. This indicates that
\begin{equation}
    - \langle p^k, x \rangle =- \langle \mathcal{K}^\top \lambda^k, x  \rangle= \langle  \lambda^k, -\mathcal{K} x  \rangle
\end{equation}
This implies that $\mathcal{B}_{J}^{p^k}(x, x^k) + \frac{\alpha}{2} \| \mathcal{K}x - f \|_2^2$ up to a constant is equivalent to the augmented Lagrangian at $\lambda^k$ as follows:
\begin{equation}
    \mathcal{L}_\alpha(x, \lambda^k) = J(x) + \langle \lambda^k, f - \mathcal{K}x \rangle + \frac{\alpha}{2} \| f - \mathcal{K}x \|_2^2. 
\end{equation}
This leads to an equivalent update of $x^{k+1}$ in (\ref{update B}) as
\begin{align}\nonumber \label{minbregman}
    &x^{k+1} = \arg\min_x \mathcal{L}_\alpha(x, \lambda^k),\\
    &\lambda^{k+1} = \lambda^k + \alpha (f - \mathcal{K}x^{k+1})
\end{align} 
\subsection{Equivalence to ADMM}
Bregman iteration is closely related to the Alternating Direction Method of Multipliers (ADMM). In fact, Esser \cite{esser2009applications} demonstrated that the Bregman framework applied to the optimization problem (\ref{problem bregman}) is equivalent to ADMM in the case of linear constraints. 
The Bregman iteration method exerts an alternating minimization approach to minimize (\ref{minbregman}) by iterating 
\begin{align} \label{bregman ADMM}
    x^{k+1} = \arg\min_x \mathcal{L}_\alpha(x, \lambda^k),
\end{align}
$T$ times, followed by updating
\begin{equation}
    \lambda^{k+1} = \lambda^k + \alpha (f - \mathcal{K}x^{k+1}).
\end{equation}

As a result, this procedure can be viewed as alternately minimizing the augmented Lagrangian $\mathcal{L}_\alpha(x, \lambda^k)$ with respect to $x$ and updating the Lagrange multiplier $\lambda$. When $T=1$, it reduces to ADMM.
This equivalence highlights the flexibility of the Bregman iteration framework, as it provides an alternative interpretation of ADMM, particularly in scenarios involving constrained optimization.
\section{Problem formulation}\label{sec2}

\subsection{Pose graph optimization}\label{subsec2}

As explained in the previous section, pose graph optimization (PGO) estimates unknown poses from the noisy relative pose measurements and we intend to jointly approximate the trajectories of all robots with regard to these relative measurements in a global coordinate frame. Let $\mathcal{V}=\{1, ..., n\}$ denote the set of $n$ poses in which each pose can be either the pose of multiple robots moved over time (like multi-robot SLAM). Accordingly, the PGO problem can be simply modeled as a directed graph, in which poses are represented by the nodes $\mathcal{V}$ and the measurements among poses are denoted by the edges $\mathcal{E \subset \mathcal{V} \times \mathcal{V}}$ in a graph. Hence a graph $G=(\mathcal{V},\mathcal{E})$ is built. In a pose graph, $m := |\mathcal{E}|$ and $n := |\mathcal{V}|$ are the number of edges and nodes, respectively.

A pose consists of its translation $\mathbf{t}_i \in \mathbb{R}^n$ and its rotation $\R_i \in \mbox{SO}(n)$, where the group $\mbox{SO}(n)$ is defined as $\mbox{SO}(n)=\{\R \in \mathbb{R}^{n \times n}: \R^\top \R=I_n, \mbox{det}(\R)=1\}$. The relative poses of node $i$ will be defined by $x_i$ and thereby $x=[x_1, x_2, ..., x_n] \in \mbox{SE(n)}$ represents the poses that need to be estimated. The special Euclidean group $\mbox{SE(n)}$ is described as \cite{alcocer2006estimation}
\begin{equation}
    \mbox{SE}(n)= \left\{ \begin{bmatrix} \R & \mathbf{t} \\ 0 & 1\end{bmatrix}: \R \in \mbox{SO}(n), \mathbf{t} \in \mathbb{R}^n \right\}.
\end{equation}
\begin{figure}[t!]
\center
\includegraphics[scale=0.75]{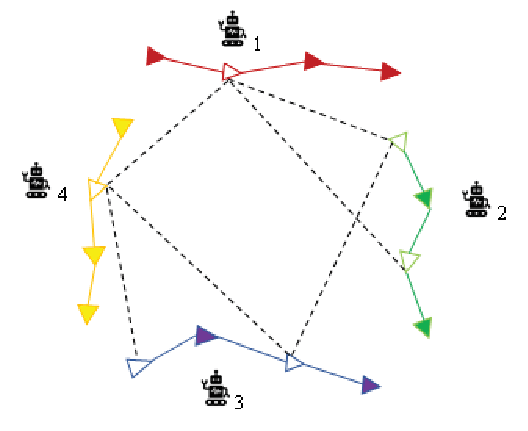}
\caption{Pose graph structure instance with 4 robots, each with 4 poses. Each edge represents a relative pose measurement. The inter-robot measurements are illustrated with the dotted line between two connected robots. The solid line denotes the intra-robot measurements of each robot.   }
\label{posegraph}
\end{figure}
The edge $(i,j)\in\mathcal{E}$ illustrates the availability of the relative pose measurements between agent $i$ and $j$, which is denoted by $x_{ij}$. The relative pose measurements, which include the relative translation $\mathbf{t}_{ij}$ and relative rotation $\R_{ij}$, are given by \cite{carlone2015initialization}:
\begin{equation}
    x_{ij}: \begin{cases}
      \mathbf{t}_{ij}=\R_i^\top  (\mathbf{t}_j - \mathbf{t}_i)+\mathbf{t}_{ij}^\epsilon\\
      \R_{ij}=\R_i^\top \R_j \R_{ij}^\epsilon
     \end{cases}
 \end{equation}
where $\mathbf{t}_{ij}^ \epsilon$ and $\R_{ij}^ \epsilon$ denote the measurement noise on the translation and the rotation, respectively. It is assumed that the noisy measurements have been acquired prior to solving the problem, unlike in visual odometry (VO) or online simultaneous localization and mapping (SLAM) framework. Two kinds of measurements are considered: intra-robot and inter-robot measurements. The intra-robot measurements are concerned with the robot's positions at different points in time which are illustrated by the solid lines in Fig \ref{posegraph}. The inter-robot measurements are obtained from the relative positions of different robots. Note that a robot with a local frame of \textit{i} may measure the position of another robot with a local frame of \textit{j} using its own sensors. This results in an inter-robot measurement, which captures the relative position of the observed robot in the reference frame of the observing robot, as depicted by dashed lines in Fig \ref{posegraph}. 

The objective of the pose graph optimization problem is to estimate the poses (consisting of translations and rotations) by minimizing the mismatch between the poses $x_i$ and the measurements $x_{ij}, \forall (i,j)\in\mathcal{E}$ given by \cite{carlone2015initialization}:
\begin{equation} \label{min} 
    \min_{\substack{\R_i \in \mbox{SO}(n)\\ \mathbf{t}_i \in \mathbb{R}^n}} \sum_{(i,j) \in \mathcal{E}} \mbox{dist}_{\mathbb{R}^n} (\mathbf{t}_{ij}, \R_i^\top(\mathbf{t}_j-\mathbf{t}_i))^2
    + \mbox{dist}_{\small{\mbox{SO}(n)}} (\R_{ij},\R_i^\top \R_j)^2.
\end{equation}

The requirement that the matrices $\R_i$ satisfy $\R_i \in \mbox{SO}(n)$ means that any matrix which minimizes (\ref{min}) must be a rotation matrix. However, solving (\ref{min}) is difficult since the constraint $\R_i\in\mbox{SO}(n)$ makes the optimization problem non-convex. The function $\mbox{dist}_{\mathbb{R}^n}(\cdot)$ denotes the Euclidean distance
\begin{equation} \label{translation}
    \mbox{dist}_{\mathbb{R}^n}  (\mathbf{t}_{ij}, \R_i^\top (\mathbf{t}_j-\mathbf{t}_i))=\norm{ \R_i^\top (\mathbf{t}_j-\mathbf{t}_i) - \mathbf{t}_{ij}}_2 =\norm{\mathbf{t}_j-\mathbf{t}_i - \R_i \mathbf{t}_{ij}}_2
\end{equation}
and the cordal distance is applied for $\mbox{dist}_{\mbox{SO}(n)}(\cdot)$
\begin{equation} \label{rotation}
    \mbox{dist}_{\mbox{SO}(n)}  (\R_{ij},\R_i^\top \R_j)=\norm{ \R_{ij}^\top \R_i^\top - \R_j^\top }_F.
\end{equation}
where $\|.\|_F$ denotes the Frobenius norm. So, by substituting (\ref{translation}) and (\ref{rotation}) into (\ref{min}), we obtain the following minimization problem

\begin{equation} \label{min overall}
    \min_{\substack{\R_i \in \mbox{SO}(n)\\ \mathbf{t}_i \in \mathbb{R}^n}} \sum_{(i,j) \in \mathcal{E}} \norm{\mathbf{t}_j-\mathbf{t}_i - \R_i \mathbf{t}_{ij}}_2^2 + \norm{ \R_{ij}^\top \R_i^\top - \R_j^\top }_F^2.
\end{equation}

Equation (\ref{min overall}) demonstrates that the objective is a quadratic function. The problem's complexity stems from the fact that the constraint $\R_i \in \mbox{SO}(n)$ is non-convex. The optimization problem (\ref{min overall}) can be solved in a centralized strategy by using convex relaxations \cite{rosen2019se}, fast approximation \cite{carlone2015initialization}, and iterative optimization \cite{dellaert2012factor} on the manifold. However, in this paper, we attempt to solve the problem in a distributed manner. For the intention of designing distributed algorithms, two types of measurements are considered: the relative measurements contain the robot $i$'s own trajectory and the relative measurements among any of robot $i$'s and $j$'s poses, as shown in Fig \ref{posegraph}. For instance, in the scenario shown in Fig \ref{posegraph}, robot 1  contains intra-measurements and just two inter-measurements with its neighboring poses in robot 2's trajectory. 

\subsection{The SOC Algorithm}

The existence of nonconvexity in distributed PGO problems increases the computational complexity of distributed optimization. To tackle this problem, the distributed method of splitting orthogonality constraints (SOC) \cite{lai2014splitting} is used. The SOC method employs variable splitting and Bregman iteration \cite{osher2005iterative, yin2008bregman} in order to solve problems constrained by orthogonality. Our algorithm consists of two layers. In the first stage, we use the SOC method for nonlinear equality-constrained problems as proposed in \cite{lai2014splitting} and solve non-convex subproblems of rotation $\R_i$. In addition, the determinant of the rotation constraint is applied in each iteration to obtain the rotation matrix. Then, any factor graph optimizer, such as the Gauss-Newton method, can be applied for the minimization in (\ref{min overall}). It is noteworthy that our proposed method converges to solutions near the global minimum. 

\begin{algorithm}[t]
\caption{SOC Algorithm}\label{algo1: SOC}
\begin{algorithmic}[1]
\State Initialization $\R_i^0 , \omega_i^0= \R_i^0, B_i^0=0, \rho>0$, and $k=1$ 
\While{stopping criteria is not met}
            \State $\R_i^k= \operatorname*{argmin}_{\R_i^{k-1}} \sum_{(i,j) \in \mathcal{E}} \| \R_{ij}^{(k-1)\top} \R_i^{(k-1)\top} - \R_j^{(k-1)\top} \|_F^2 + \frac{\rho}{2} \| \R_i^{k-1}- \omega_i^{k-1} +B^{k-1} \|_F^2$,
            \State Let $Y_i^k= \R_i^k +B_i^{k-1}$.  Compute SVD factorization $Y^k=UDV^\top$
            \State $\omega_i^k=U I_{n\times n} V^\top$.
            \State $B_i^k=B_i^{k-1} + \R_i^k -\omega_i^k$.
            \State $k \leftarrow k+1$.
\EndWhile
\State  Output $\R_i$ and $\omega_i$
\end{algorithmic}
\end{algorithm}

Consider the following rotation subproblem:
\begin{equation} \label{rotation subproblem with constraint}
     \min_{\R_i \in \mbox{SO}(3)} \sum_{(i,j) \in \mathcal{E}} \| \R_{ij}^\top \R_i^\top - \R_j^\top \|_F^2
\end{equation}

The above rotation subproblem can be reformulated by introducing an auxiliary variable $\omega_i \in \mathbb{R}^{3 \times 3}$ to split the orthogonality constraints as follows:
\begin{equation} \label{relaxation rotation}
    \begin{split}
     & \min_{\R_i, \omega_i \in \mathbb{R}^{3\times 3}} \sum_{(i,j) \in \mathcal{E}} \| \R_{ij}^\top \R_i^\top - \R_j^\top \|_F^2\\
     & S.t. \hspace{0.5cm} \omega_i^\top \omega_i=I,\hspace{0.35cm} \omega_i=\R_i.
    \end{split}
\end{equation}

\begin{figure}
    \centering
    \includegraphics[trim = 1mm 30mm 1mm 40mm, width=\textwidth]{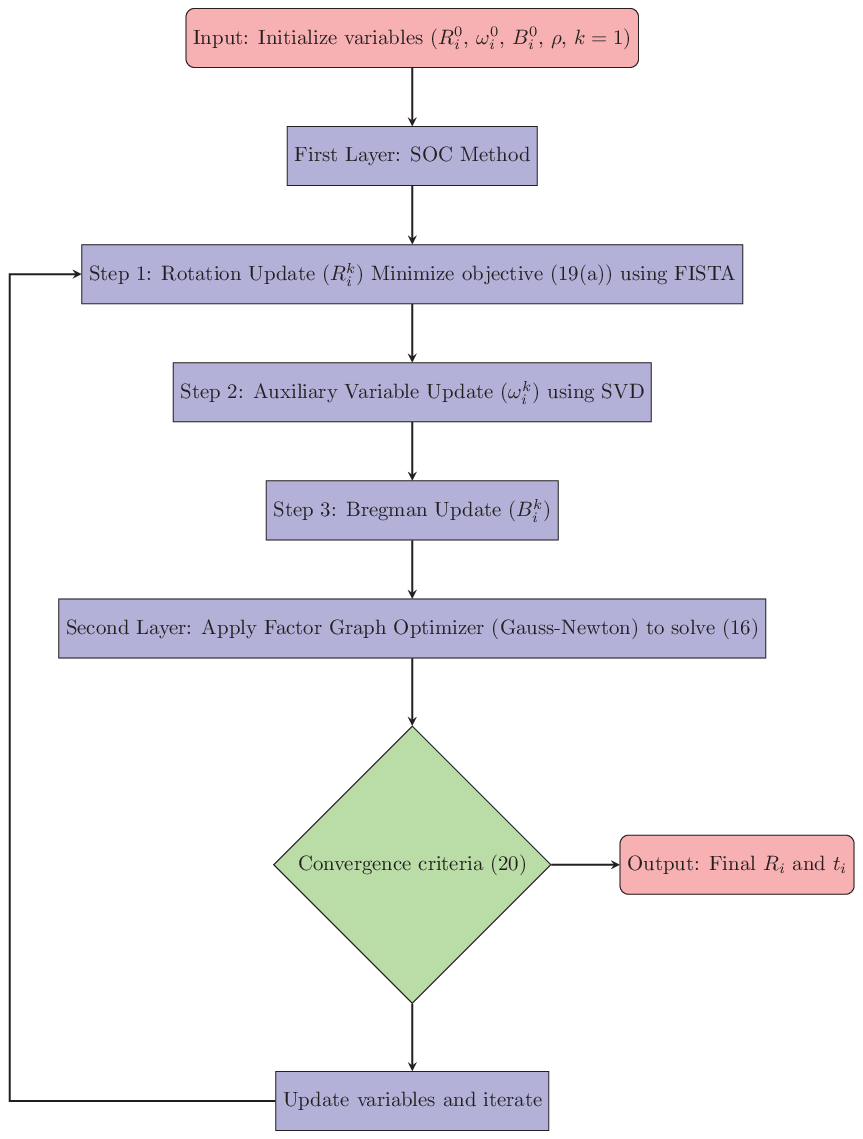}
    \caption{The flow chart of the proposed algorithm.}
    \label{flowchart}
\end{figure}

In this case, problem (\ref{relaxation rotation}) is indicated as the orthogonal relaxation of rotation subproblem, acquired by dropping the determinant constraint on $\mbox{SO}(3)$. To address the minimization problem described in (\ref{relaxation rotation}), the Splitting Orthogonality Constraints (SOC) method can be employed, which updates variables $\omega_i$ and $\R_i$ in an alternating fashion by utilizing the concepts of the Alternating Direction Method of Multipliers (ADMM) and the Split Bregman algorithm. Therefore, the problem can be solved iteratively through the use of Bregman iteration, as demonstrated below: 
\begin{equation} 
\begin{cases}
     (a)\hspace{0.1cm} \R_i^k= \operatorname*{argmin}_{\R_i^{k-1}} \sum_{(i,j) \in \mathcal{E}} \| \R_{ij}^{(k-1)\top} \R_i^{(k-1)\top} - \R_j^{(k-1)\top} \|_F^2  \\ \hspace{4.5cm} +  \frac{\rho}{2} \| \R_i^{k-1}- \omega_i^{k-1} +B_i^{k-1} \|_F^2, \\ 
     (b)\hspace{0.1cm}\omega_i^k= \operatorname*{argmin}_{\omega_i^{k-1}} \frac{\rho}{2} \| \omega_i^{k-1} - (\R_i^k +B_i^{k-1})\|_F^2, \hspace{0.2cm}  \hspace{0.8cm} \text{s.t.} \hspace{0.3cm} \omega_i^\top \omega_i=I, \\ 
     (c)\hspace{0.1cm} B_i^k=B_i^{k-1} + \R_i^k -\omega_i^k.       
\end{cases}
\end{equation}
Here $\rho$ is the penalty parameter. The first subproblem is a convex optimization problem and can be resolved by employing the Fast Iterative Shrinkage-Thresholding Algorithm (FISTA) \cite{beck2009fast}. The second subproblem has a closed-form solution according to \cite[Theorem 1]{lai2014splitting}). Algorithm \ref{algo1: SOC} illustrates the procedure of updating the variables alternately. A flow chart illustrating the structure and key steps of the proposed algorithm is depicted in Fig. \ref{flowchart}, to enhance understanding of the methodology.

Note that the obtained rotation projects to $\mbox{SO}(3)$ by performing SVD are accomplished independently for each iteration.  
In contrast to other distributed methods \cite{choudhary2017distributed, cristofalo2019consensus}, our techniques contain the subtlest conditions (\textit{e.g.} no assumptions on the pose graph structures are made and any master node for information accumulation or preprocessing can be chosen) to converge closely to the optimal solution. Lastly, the rotation and translation approximation can be implemented concurrently by using the Gauss-Newton method for the minimization problem in (\ref{min overall}). 

A standard method to evaluate the convergence of optimization problems is to verify the Karush-Kuhn-Tucker (KKT) conditions. The KKT conditions are exerted to (\ref{relaxation rotation}). The first condition, which is entitled primal feasibility is for satisfying the constraints. This stopping condition is defined in \cite{choudhary2015exactly}. The stationary condition satisfies the requirement of the solution to be a stationary point of \eqref{relaxation rotation}. Therefore, the algorithm is terminated when the values of the residual are met by assuming the desired accuracy: 
\begin{equation}
    p_{res} \leq \sqrt{ \norm{ (\omega^{{k+1})^\top} \omega^{k+1} -I}_F^2 + \norm{\omega^{k+1} -\R^{k+1} }_F^2}.
\end{equation}

A practical matter in executing our algorithm to be appropriate for solving distributed PGO is how to determine the value of the penalty parameter $\rho$. Setting $\rho$ to a constant value would result in slow convergence of the algorithm. Hence, an adaptive scheme from \cite{boyd2011distributed} is considered. The main point is that if the primal residual is large at iteration $k$, then a greater $\rho$ will be adopted at the next iteration. However, if the dual residual is large, $\rho$ will be decreased as attained the stationary point of the Lagrangian. Hence, the same penalty parameter update rule presented in \cite{choudhary2015exactly} is adopted.
\begin{remark} (The distinct features of the SOC algorithm)
   In this study, we address the nonconvex rotation subproblem using the SOC method. To the best of our knowledge, this is the first article to leverage the SOC approach for solving the distributed PGO problem. To implement this method, we employ the ADMM algorithm and the split Bregman iteration technique. By employing the SOC approach, we can effectively handle spherical constrained problems through a sequential process that iteratively minimizes an unconstrained problem and a quadratic problem incorporating orthogonality constraints. Of particular significance is the fact that the solution to the quadratic problem, considering the orthogonality constraints, can be analytically expressed as a spherical projection. This provides an efficient methodology for effectively managing the challenges presented by orthogonality constraints in the given problem. These distinct characteristics of the SOC method differentiate it from penalty methods and standard augmented Lagrangian methods. Penalty methods experience slow convergence as the penalty parameter approaches zero, due to the need to address a series of problems, while augmented Lagrangian methods introduce computational overhead from additional inner iterations to navigate the augmented Lagrangian terms, further hampering convergence speed. 
\end{remark}
\begin{remark}
    One concrete application of our algorithm is in multi-robot SLAM, where robots collaboratively estimate their poses while mapping an environment. For instance, consider a scenario where a fleet of robots equipped with sensors (e.g., cameras, LiDAR) must localize themselves within a shared environment. The robots share relative pose measurements, which are inherently noisy, and the task is to optimize the pose estimates across all robots while respecting the non-convex orthogonality constraints of rotation matrices. Furthermore, a Sparse Low-Rank Matrix Approximation (SLRMA) method for data compression is required to simultaneously incorporate both the intra-coherence and inter-coherence of data samples from the perspective of optimization and transformation. The SLRMA problem can be expressed as a constrained optimization problem while satisfying both the $l_0$-norm sparsity and orthogonality constraints. Our proposed method can be applied to address this optimization problem.
\end{remark}

\section{Simulation results}\label{sec3}

The evaluation of the proposed algorithm encompasses two primary criteria: accuracy and efficiency. Accuracy is quantified by the percentage error of the algorithms' results, while efficiency is gauged by considering the number of iterations and the time required to achieve convergence. These efficiency metrics serve as crucial indicators of the practical feasibility and performance of the proposed method in real-world applications. The experiment setup closely follows the experimental framework presented by Cristofalo et al. \cite{cristofalo2020geod}, ensuring consistency and comparability with previous work in the field.

\begin{figure} [t]
\captionsetup[subfigure]{justification=Centering}

\begin{subfigure}[t]{0.3\textwidth}
    \includegraphics[trim = 1mm 10mm 1mm 20mm, width=\textwidth]{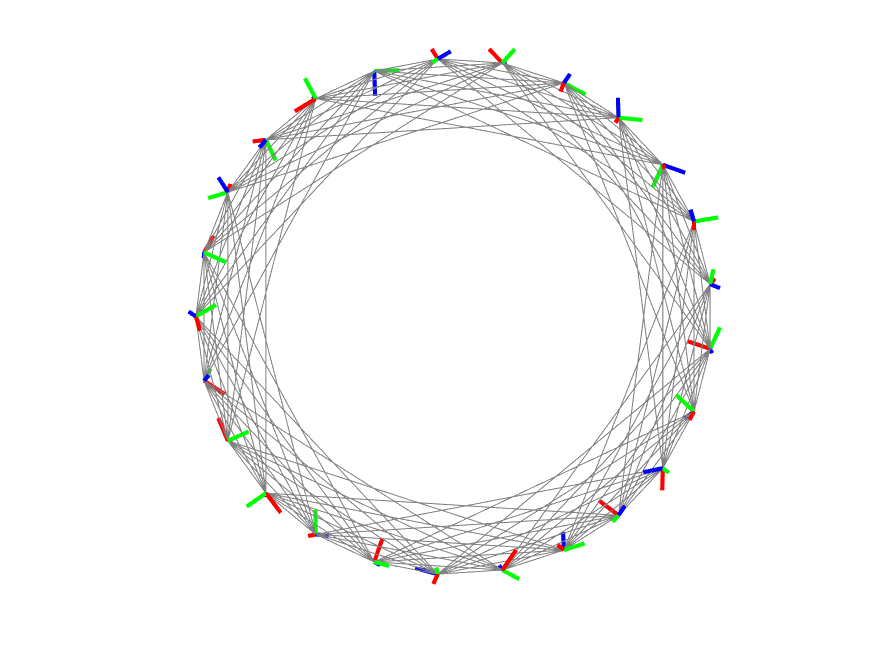}
    \caption{Circle ground truth}
\end{subfigure} 
\begin{subfigure}[t]{0.35\textwidth}
    \includegraphics[trim = 1mm 20mm 0mm 20mm, width=1.1\textwidth]{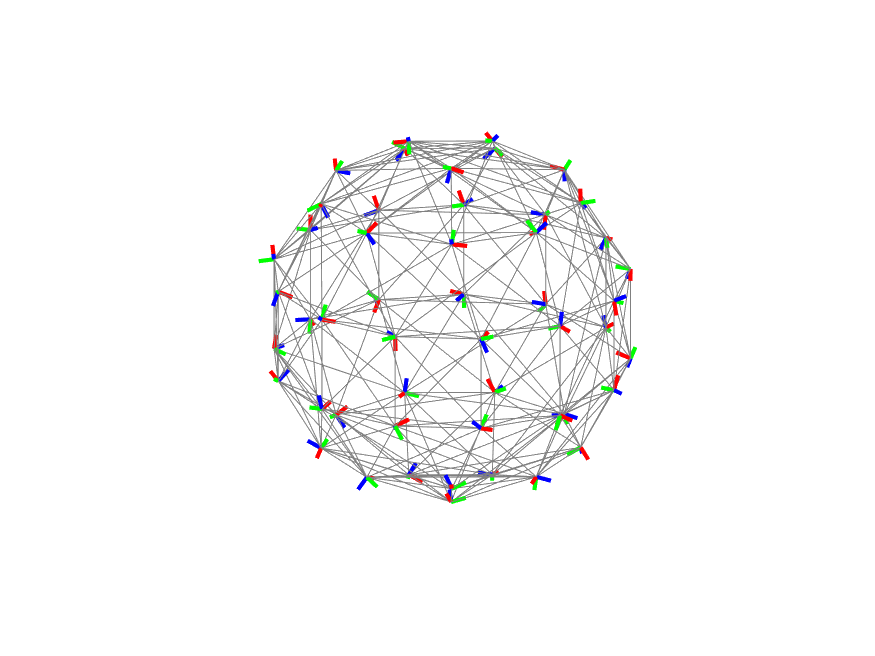}
    \caption{Sphere ground truth}
\end{subfigure}
\begin{subfigure}[t]{0.32\textwidth}
    \includegraphics[trim = 1mm 10mm 0mm 20mm, width=\linewidth]{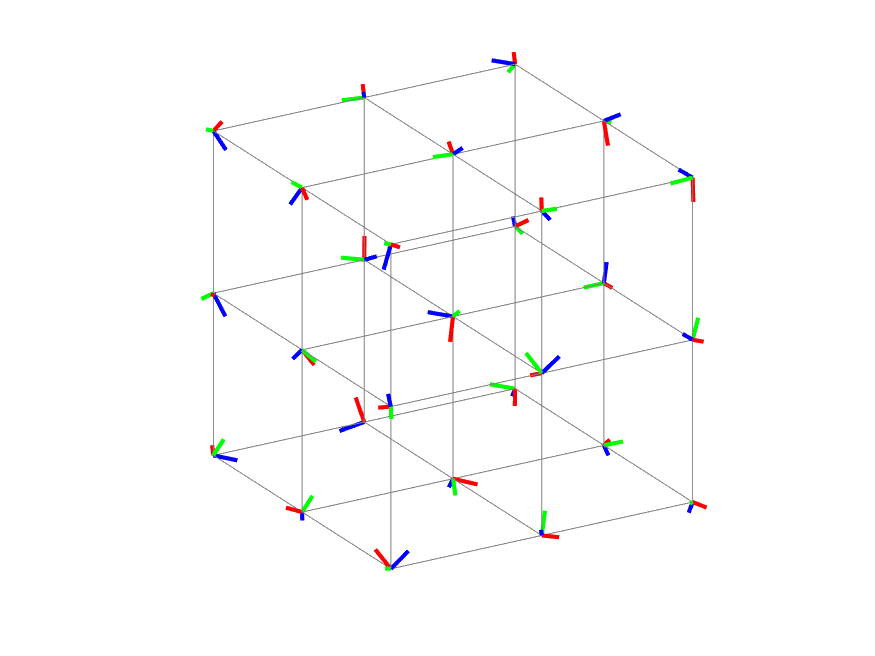}
    \caption{Grid ground truth}
\end{subfigure}

\bigskip 
\begin{subfigure}[t]{0.3\textwidth}
    \includegraphics[trim = 1mm 2mm 0mm 20mm,width=\linewidth]{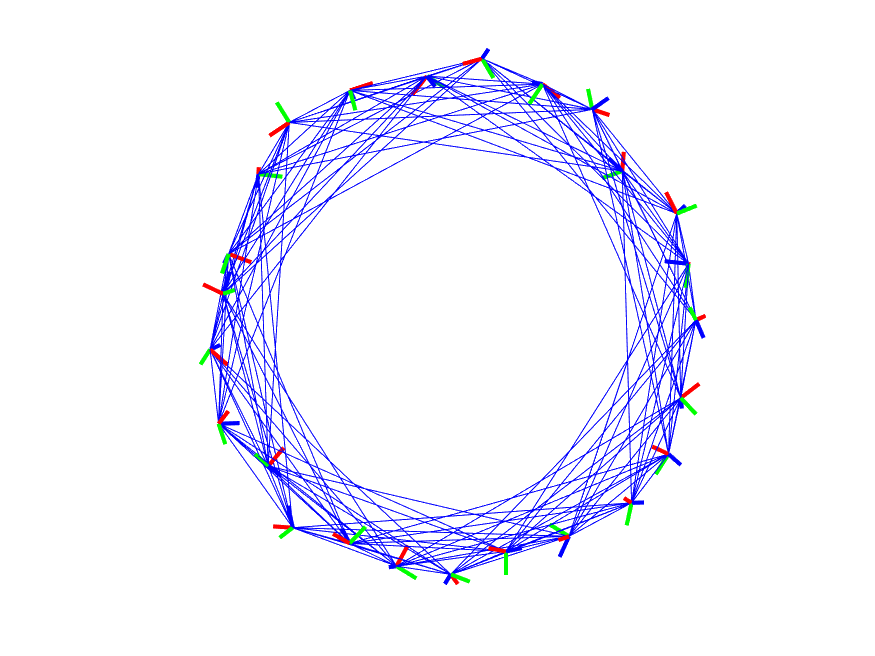}
    \caption{ SOC result}
\end{subfigure} 
\begin{subfigure}[t]{0.35\textwidth}
    \includegraphics[trim = 1mm 18mm 0mm 10mm,width=1.1\textwidth]{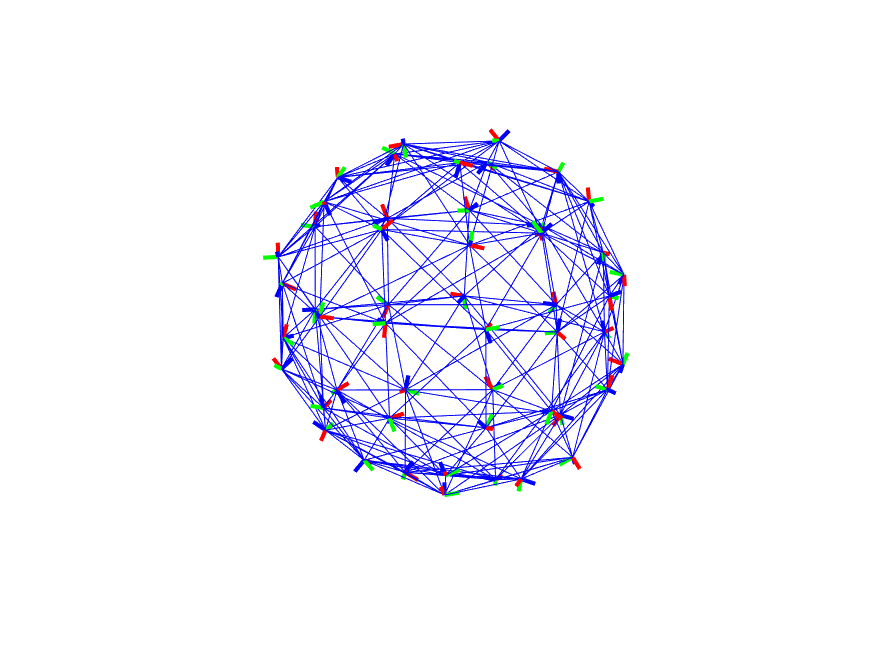}
    \caption{ SOC result}
\end{subfigure}
\begin{subfigure}[t]{0.32\linewidth}
    \includegraphics[trim = 1mm 18mm 0mm 10mm,width=1.2\linewidth]{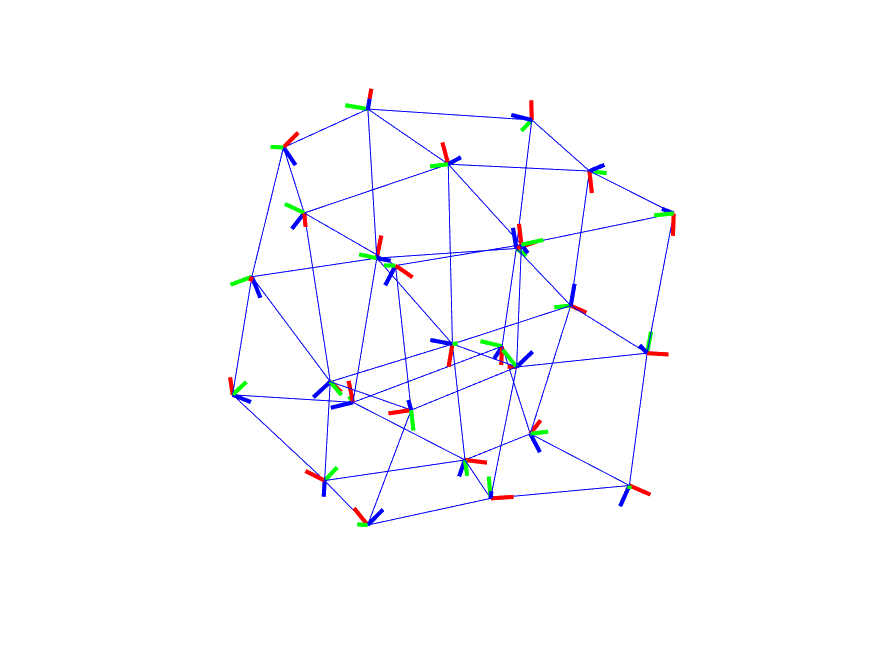}
    \caption{SOC result}
\end{subfigure}

\caption{Circular, sphere, and grid pose graphs from left to right. The first row demonstrates the ground truth of network topologies. The second row illustrates the obtained solution by applying the SOC algorithm.}
\label{figure1}

\end{figure}
\begin{figure} [t!]
\captionsetup[subfigure]{justification=Centering}

\begin{subfigure}[t]{0.3\textwidth}
    \includegraphics[trim = 1mm 10mm 1mm 20mm, width=\textwidth]{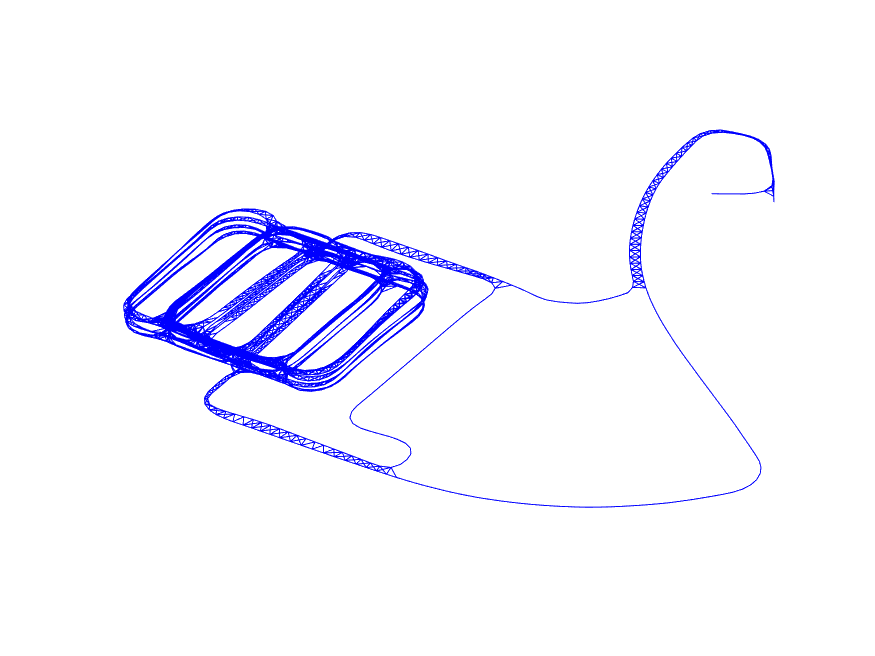}
    \caption{Parking garage's SOC result}
\end{subfigure} 
\begin{subfigure}[t]{0.35\textwidth}
    \includegraphics[trim = 1mm 20mm 0mm 20mm, width=1.1\textwidth]{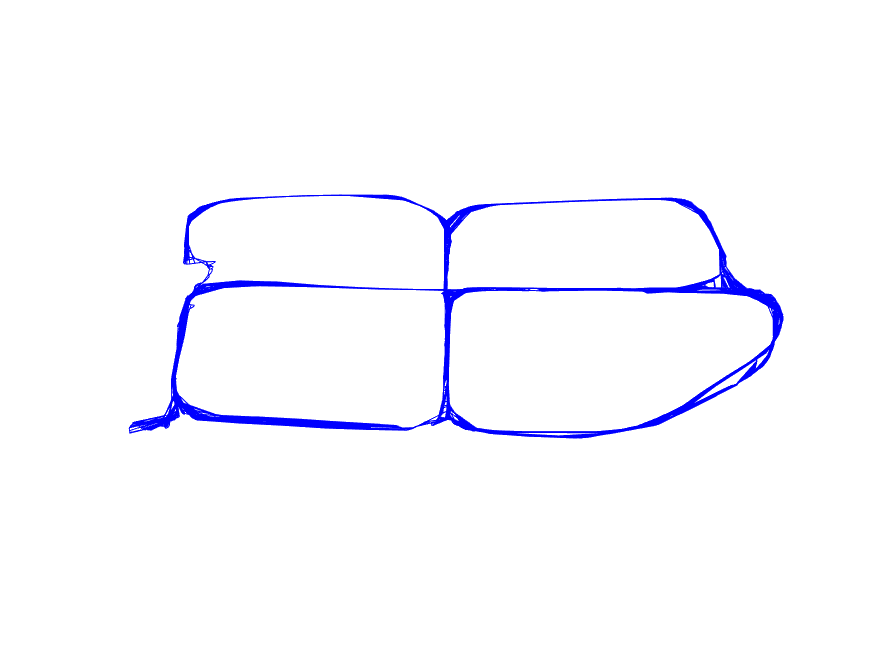}
    \caption{Cubicle's SOC result}
\end{subfigure}
\begin{subfigure}[t]{0.32\textwidth}
    \includegraphics[trim = 1mm 10mm 0mm 20mm, width=1\linewidth]{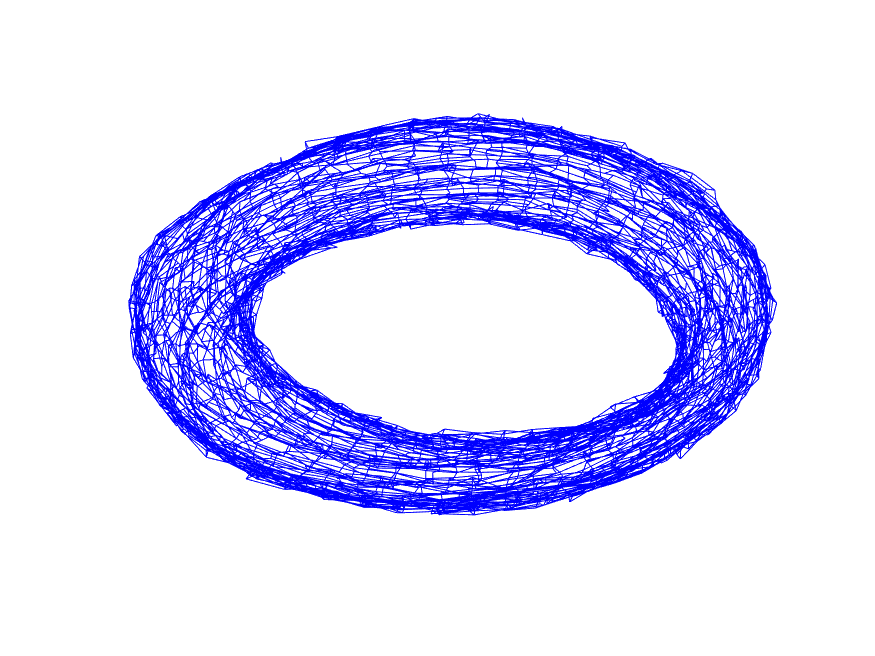}
    \caption{Torus's SOC result}
\end{subfigure}

\caption{The outcomes regarding the SOC algorithm for the parking garage, cubicle, and torus pose graphs are depicted from left to right.}
\label{figure3}
\end{figure}

In order to validate the effectiveness of the proposed algorithm, we conduct experiments on simulated and real-world pose graph datasets. The performance of our algorithm is compared against two state-of-the-art methods: the centralized Se-Sync approach \cite{rosen2019se} and the Distributed Gauss-Seidel (DGS) method \cite{choudhary2017distributed}. The Se-Sync method solves a semi-definite convex relaxation of the pose graph optimization problem to obtain a global minimum, while DGS addresses a convex chordal relaxation problem in a distributed scheme. The SOC method offers an alternative approach to address the optimization task by considering splitting orthogonality constraints.

We extensively evaluate the performance of our proposed algorithms on a diverse set of seven network topologies, including circular, grid, sphere, parking garage, cubicle, torus, and rim networks. Among these, the first three networks are simulated datasets, while the remaining four are established SLAM benchmark datasets \cite{rosen2019se}.
For the circular, grid, and sphere networks, we have access to ground truth topologies and translations, while the ground truth rotation matrices are uniformly and randomly sampled as specified in \cite{arvo1992fast}. To create pairwise relative measurements for these networks, noise is introduced to the ground truth-related poses. Specifically, translation noise is added by drawing samples from a Gaussian distribution $\mathcal{N} (0, \tau^2 I_3)$ with $\tau=0.5$. The rotation noise is generated by sampling from the vector space of SO(3) and subsequently utilizing the exponential map to return to the group. On the other hand, the parking garage, cubicle, torus, and rim networks, which are SLAM benchmark datasets, provide noisy measurements without any ground truth positioning information. These real-world datasets enable us to rigorously assess the effectiveness of our algorithms under challenging conditions and demonstrate the scalability and robustness of our approach in handling large pose graphs.

In our simulations, the pose approximation is initialized via simulated GPS measurements, except for the SLAM benchmark datasets, which are initialized by the spanning tree initialization method \cite{cristofalo2020geod} to make the comparisons
fair. The translation and rotation approximation is assumed to be provided using GPS and an Inertial Measurement Unit (IMU).

The ground truth of translation is corrupted by adding zero-mean Gaussian noise to generate the initial translation approximation as acquired from the same distribution as the aforementioned relevant measurement noise. Awerbuch et al. \cite{awerbuch1987optimal} proposed a distributed algorithm for obtaining a spanning tree that can be employed in a time in terms of the number of robots in the network. The initial rotation approximation follows a similar process to that of the relative rotation measurements and employs the same distribution: $\hat{\R}_i=\R_i \exp (d)$, with $d$ sampled from $\mathcal{N} (0, \nu^2 I_3)$, where $\nu$ is set to 30 degrees.

The circular, sphere, and grid graphs used for simulations are depicted in Figure \ref{figure1}. The first row of Figure \ref{figure1} illustrates the ground truth of the mentioned network topologies, while the second row shows the obtained results after applying the SOC algorithm.
The results obtained from employing the SOC algorithm for the parking garage, cubicle, and torus are depicted in Figure \ref{figure3}. These benchmark datasets serve as real-world examples to evaluate the performance of the algorithm in practical scenarios.
\begin{table}[t!]
\caption{The comparison of PGO objective value outcomes between the state-of-the-art centralized SE-Sync \cite{cristofalo2020geod,rosen2019se}, distributed DGS \cite{cristofalo2020geod, choudhary2017distributed}, and the proposed method (SOC). * represents the certified global minimum gained by SE-Sync}.
\begin{tabular}{cccccccc}
\toprule
& &  &   & \multicolumn{2}{c}{\makecell{$F$}} \\ 
\cmidrule{5-6} 
 Scene& \# poses& \# measurements&$F^*$ (SE-Sync) \cite{cristofalo2020geod} &   DGS  \cite{cristofalo2020geod} & SOC             \\
\midrule 
\midrule
Circle & 25  &  300   &    315.98  & 415.762 &  \textbf{328.756}         \\ 
\midrule
Grid  &  27 &  108   &   85.774    &  123.244    & \textbf{92.831}        \\ 
\midrule
Sphere  & 50  & 544    &  558.386     & 688.637     & \textbf{594.230}        \\ 
\midrule
\end{tabular}
\label{table1}
\end{table}

\begin{table}[t!]
\caption{Comparison between the run times of each method, as well as the communication iterations to convergence.}
\begin{tabular}{ccccccc}
\toprule
 &   \multicolumn{3}{c}{\makecell{Iteration}} & \multicolumn{3}{c}{\makecell{Time}}  \\ 
\cmidrule{2-4} \cmidrule{5-7} 
Scene &  SE-Sync  \cite{cristofalo2020geod}&  DGS  \cite{cristofalo2020geod}& SOC   & SE-Sync \cite{cristofalo2020geod}&  DGS \cite{cristofalo2020geod} & SOC        \\
\midrule 
\midrule
Circle & -  &  15   &    17   &   0.097   & \textbf{0.0027} & 0.007   \\ 
\midrule
Grid  &  - &  \textbf{10}   &   78     &0.059 & \textbf{0.0013}   &  0.011    \\ 
\midrule
Sphere  & -  & \textbf{10}    &  38    & 0.11  & \textbf{0.0020}  & 0.010  \\ 
\midrule
\end{tabular}
\label{table2}
\end{table}

\begin{table}[t!]
\caption{The outcomes of distributed Pose Graph Optimization (PGO) on the 3D SLAM Benchmark datasets. The objective value is denoted as $F$, while the globally optimal objective value is represented as $F^*$. }
\begin{tabular}{lcccccc}
\toprule
& &  &   & \multicolumn{2}{c}{\makecell{$F$}} \\ 
\cmidrule{5-6} 
 Scene& \# poses& \# measurements&$F^*$ (SE-Sync) \cite{cristofalo2020geod} &   DGS  \cite{cristofalo2020geod} & SOC             \\
\midrule 
\midrule
Parking \\garage &  1661 & 6275  &    1.263  & 3.801 &  \textbf{2.31}         \\ 
\midrule
Cubicle  & 5750 & 16869   &  717.126    &  -    & 724.56        \\ 
\midrule
Torus & 5000 & 9048     &  24227.046    & -     & 24449.94        \\ 
\midrule
Rim & 10195 & 29743    &  5460.89     & -    & 6182.73       \\ 
\midrule
\end{tabular}
\label{table3}
\end{table}

\begin{table}[t!]
\caption{The comparison of PGO objective computation time outcomes between centralized SE-Sync, distributed DGS, and the proposed SOC method }
\begin{tabular}{lcccccc}
\toprule
 &   \multicolumn{3}{c}{\makecell{Iteration}} & \multicolumn{3}{c}{\makecell{Time}}  \\ 
\cmidrule{2-4} \cmidrule{5-7} 
Scene &  SE-Sync  \cite{cristofalo2020geod}&  DGS  \cite{cristofalo2020geod}& SOC   & SE-Sync \cite{cristofalo2020geod}&  DGS \cite{cristofalo2020geod} & SOC        \\
\midrule 
\midrule
Parking \\garage &  - &  265  &   \textbf{98}   &   84.71   & 2.62 & \textbf{0.094}   \\ 
\midrule
Cubicle  & - & - &   4980     &23.528 & -   &  \textbf{3.11}    \\ 
\midrule
Torus & - &   -     &  3175   & 17.26  & -  & \textbf{2.35} \\ 
\midrule
Rim & - &    -   & 14352    & 41.56   & -  & \textbf{5.22}  \\
\midrule
\end{tabular}
\label{table4}
\end{table}

Table \ref{table1} presents a comprehensive comparison of the objective values obtained through the application of the three methods on simulated datasets (circle, grid, and sphere). In Table \ref{table2}, we contrast the execution times of each method and the number of communication iterations required to achieve convergence. Additionally, the error percentage is calculated by comparing the objective value with the certified global minimum value. Notably, SE-Sync, operating as a centralized method, does not involve communication iterations and successfully achieves the certified global minimum in the scenarios considered.

Referring to Table \ref{table2}, our proposed SOC method demonstrates convergence with an average error rate of 6.22\% across the first three simulated datasets (circle, grid, and sphere). Conversely, the DGS algorithm achieves convergence with an error rate of 32.96\%. As indicated in Table \ref{table3}, our method showcases significant convergence performance, attaining an error rate of more than 50\% lower than that of the DGS approach in the case of the parking garage dataset. Concerning computation time, our algorithm surpasses the centralized SE-Sync method in terms of speed. However, in the top three datasets, the DGS approach demonstrates faster execution, whereas it falls behind in the parking garage dataset.

The experimental outcomes of the real-world SLAM dataset are depicted in Table \ref{table3} and \ref{table4}, wherein $F^*$ represents the objective value achieved by SE-Sync \cite{rosen2019se}, denoting the globally optimal objective. On the other hand, $F$ denotes the objective value attained by each respective method.
The run times of each method and the number of communication iterations to attain convergence are compared in Table \ref{table4}. It should be noted that SE-Sync is a centralized method, so there are no communication iterations. Moreover, SE-Sync achieves the certified global minimum in considered scenarios.
As shown in Table \ref{table3}, the proposed method outperforms the DGS method \cite{choudhary2017distributed} for the parking garage dataset. While Table \ref{table4} details run times and communication iterations for the three methods in SLAM benchmark datasets, it's important to note that this information is not available for the DGS algorithm in \cite{cristofalo2020geod}, limiting direct comparisons. Nonetheless, in terms of run time, the SOC method consistently outperforms SE-Sync across all SLAM benchmark datasets.

It is worth noting that as the number of agents increases, the DGS algorithm encounters convergence issues and fails to yield results for cubicle, torus, and rim datasets, as highlighted in Table \ref{table3}. In contrast, our proposed method remains applicable to graphs with a substantial number of poses. While our method may not attain the global minimum, it consistently approaches near-global minima, as evidenced in Table \ref{table1} and Table \ref{table3}. Importantly, our method significantly outperforms the DGS algorithm in scenarios involving extensive networks.

\begin{remark}(Comparison with other state-of-the-art methods)
    In this study, through thorough experimentation on diverse simulated and real-world pose graph datasets, we compare our method against state-of-the-art approaches such as the centralized Se-Sync approach \cite{rosen2019se} and the DGS method \cite{choudhary2017distributed}. The SOC method demonstrates its clear superiority in terms of accuracy and efficiency. While the Se-Sync method relies on semi-definite convex relaxation and the DGS method addresses convex chordal relaxation, the SOC method introduces a novel pathway through orthogonality constraints. Notably, the SOC method achieves convergence with an average error rate of 6.22\% in simulated datasets, significantly outperforming the DGS method with a convergence error rate of 32.96\%. The SOC method's performance is further highlighted in real-world scenarios, where it exhibits convergence rates surpassing 50\% lower than the DGS approach. Additionally, the SOC method showcases commendable computation speed, surpassing the centralized Se-Sync method in various scenarios. Although the DGS approach demonstrates faster execution in certain cases, the SOC method consistently excels in terms of run time across all SLAM benchmark datasets.  
\end{remark}

\section{Conclusion}

We have proposed a distributed non-convex algorithm to pose graph optimization problems while illustrating convergence. The proposed approach provides the following merits: \textit{i)} the complex non-convex rotation subproblem is effectively handled through the application of a splitting technique that leverages both the alternating direction method of multipliers (ADMM) and the Bregman iteration method, \textit{ii)} our method stands out by providing solutions that are not only more accurate than the Distributed Gauss-Seidel (DGS) method but are also achieved in considerably less time compared to the centralized SE-Sync method, \textit{iii)} the simplicity of our proposed approach enhances its practical applicability, making it accessible to a broader range of researchers in the field. Through comprehensive comparisons with state-of-the-art techniques, we thoroughly evaluate our method's performance in terms of both accuracy and efficiency. This comprehensive analysis provides the advantages of our proposed approach, underscoring its potential to significantly contribute to the advancements in distributed non-convex pose graph optimization.

\backmatter

\bmhead{Acknowledgements}
Not applicable.

\section*{Declarations}

\begin{itemize}
\item Funding: The authors did not receive support from any organization for the submitted work.
\item Conflict of interest/Competing interests: The authors declare no conﬂicts of interest.
\item Ethics approval and consent to participate: Research does not involve Human Participants and/or Animals.
\item Consent for publication: Yes
\item Data availability: Not applicable
\item Materials availability: Not applicable
\item Code availability: Not applicable
\item Author contribution: Zeinab Ebrahimi contributes to the algorithm design, simulation, and paper writing. Mohammad Deghat contributes to the development of the research methodology, the revision, and refinement of the manuscript. This includes making substantial edits for clarity, organization, and scientific accuracy.
\end{itemize}


\bibliography{sn-article}






\end{document}